\documentclass{pasa}

\usepackage{graphicx}

\title[Spin of BH mergers]{Spin of LIGO/VIRGO merging black holes as the result of binary evolution}

\author[Bogomazov et al.]{A. I. Bogomazov$^1$\thanks{a78b@yandex.ru}, V. M. Lipunov$^1$, A. V. Tutukov$^2$ and A. M. Cherepashchuk$^{1}$
\affil{$^1$M. V. Lomonosov Moscow State University, P. K. Sternberg Astronomical Institute, Universitetskij prospect 13, 119991, Moscow, Russia}
\affil{$^2$Institute of astronomy, Russian Academy of Sciences, Pyatnitskaya ulitsa 48, 119017, Moscow, Russia}
}

\jid{PASA}
\doi{xx.xxxx/pas.\the\year.xxx}
\jyear{\the\year}

\usepackage{aas_macros}
\usepackage{hyperref} 
\hypersetup{colorlinks,citecolor=blue,linkcolor=blue,urlcolor=blue}

\begin{document}

\begin{frontmatter}
\maketitle

\begin{abstract}
Recently discovered bursts of gravitational waves provide a good opportunity to verify the current view on the evolution of close binary stars. Modern population synthesis computer programs help to study this evolution from two main sequence stars up to the formation of compact remnants. To calculate the evolution of predecessors of black hole (BH) mergers we used the ``Scenario Machine'' code. The scenario modelling allowed to describe the evolution of systems for which the final stage is a BH+BH merger and showed the plausibility of modern evolutionary scenarios for binary stars and population synthesis models based on it. We discuss possible values of spins in BH mergers, and give arguments in favour of different values of spins of BH components in BH mergers (low spin + low spin, low spin + high spin, high spin + high spin). We predict that $\geq 50\%$ BH mergers originated through isolated binary evolution should possess spins of both BHs close to zero (low spin + low spin). Up to  $\approx$10\% of BH mergers are able to consist of two BHs with spins close to one (high spin + high spin), predecessors of such binaries could be sources of two gamma ray bursts.  The BH with the spin close to one could be formed as the result of the merger of two BHs formed via the collapse of a fast rotating star accompanied with a long gamma ray burst and/or a gravitational wave burst with smaller power in comparison with the merger of two BH remnants of binary components.
\end{abstract}

\end{frontmatter}

\section{INTRODUCTION}
\label{sec:intro}

The merger rate of compact remnants of the evolution of binary stars was studied in numerous investigations, e.g. \citet{lipunov1987,phinney1991,tutukov1993,tutukov1994,lipunov1997b, lipunov1997,lipunov1997c,nutzman2004,mink2015}.
At the moment of writing this manuscript there are five reliably detected mergers of binary black holes and one merger of a binary neutron star (GW170817, \citealp{abbott2017d}). Data about masses and spins of black hole mergers are collected in Table \ref{table1}. Low BH spin values in four of five detected mergers and BH masses in mergers higher than BH masses in high mass X-ray binaries inspire searches of alternative ways of black hole mergers origin, e.g. mergers of primordial black holes \citep{blinnikov2016}, a dynamical evolution of multiple stars \citep{antonini2016}, a chemically homogeneous evolution \citep{mink2016}, dynamical processes in star clusters \citep{rodriquez2016}, selection effects and the accretion of gas and dust in a molecular cloud through which the binary black hole travels \citep{tutukov2017}. For the isolated binary evolution scenarios the low metallicity of preceding stars in binary evolution model was proposed \citep{woosley2016,belczynski2016,tutukov2017}.

The aim of this investigation is to study a possible distribution of merging black holes (originated in the course of the isolated binary evolution) on their spins. For this purpose we review the rotational evolution of stellar nuclei and probable rotational rates of final remnants of this evolution: white dwarfs (WDs), neutron stars (NSs) and black holes (BHs). It was understood long time ago that the formation of a dense nucleus inside an evolving star may lead to a rapidly rotating final remnant, and up to now it remains uncertain whether they can get critical rotation that can lead them to split and to radiate gravitational waves \citep{tutukov1969,dorazio2018}. The latter scenario for BHs is under discussion until now \citep{postnov2017,qin2018}. The contraction of the stellar core in the course of its evolution up to a factor $\sim 10^5$ for nuclei forming BHs has to lead to a large acceleration of their rotation according to the rotational momentum conservation law. But non-avoidable ties of the contracting core with the expanding envelope should break down the rotation of the contracting core.  The efficiency of such braking force remains unclear. In this paper we discussed several simple observational and theoretical estimates about the efficiency of the core braking by the expanding envelope.

\begin{table*}
\caption{Registered BH+BH mergers. $M_{BH1}$ and $M_{BH2}$ are masses of merging BHs in $M_{\odot}$, $K_1$ and $K_2$ are their dimensionless spin magnitudes, $\chi_{eff}$ is an effective inspiral spin parameter.}
\centering
\begin{tabular*}{\textwidth}{@{}l\x l\x l\x l\x l\x l\x l@{}}
\hline\hline
Name & $M_{BH1}$ & $M_{BH2}$ & $K_1$ & $K_2$ & $\chi_{eff}$ & References\\
\hline
GW150914 & $36^{+5}_{-4}$ & $29\pm 4$ & $0.32^{+0.49}_{-0.29}$ & $0.44^{+0.5}_{-0.4}$ & $-0.07^{+0.16}_{-0.17}$ & \citet{abbott2016a,abbott2016b}\\
GW151226 & $14.2^{+8.3}_{-3.7}$ & $7.5\pm 2.3$ & ?$^a$ & ?$^a$ & $\approx 0.22$ & \citet{abbott2016c} \\ 
GW170104 & $31.2^{+8.4}_{-6}$ & $19.4^{+5.3}_{-5.9}$ & ? & ? & $-0.12^{+0.21}_{-0.3}$ & \citet{abbott2017a} \\
GW170608 & $12^{+7}_{-2}$ & $7\pm 2$ & ? & ? & $0.07^{+0.23}_{-0.09}$ & \citet{abbott2017b} \\ 
GW170814 & $30.5^{+5.7}_{-3}$ & $25.3^{2.8}_{-4.2}$ & ? & ? & $0.06\pm 0.12$ & \citet{abbott2017c} \\ 
\hline\hline
\end{tabular*}
\label{table1}
\tabnote{$^a$At least one of the black holes before merger had spin greater than 0.2 \citep{abbott2016c}.}
\end{table*} 

\section{ROTATION OF FINAL REMNANTS OF SINGLE STARS}
\label{sec:remnant-rotation}

The evolution of stars is well studied from observational and theoretical points of view \citep{masevich1988,biskogan2011,cherepashchuk2013b,biben2013}. In a simple general scheme it looks as following. The evolution of dense rotating clouds of gas and dust in the course of the collapse leads to the formation of stars with masses $0.1-150 M_{\odot}$. The birth frequency of young stars in the Milky Way galaxy is $d\nu = M^{-2.35}dM$ per year, where $M$ is the mass in $M_{\odot}$. The evolution of single stars and components of wide binaries with masses $0.8-10 M_{\odot}$ ends as degenerate dwarfs. Stars with masses $10-25M_{\odot}$ give NSs as remnants. More massive stars produce BHs (however, see \citet{sukhbold2016}, who showed that BHs may form from less massive stars, and NSs can form from more massive stars).

An analysis of observed rotational periods of $\approx 12000$ main-sequence stars in the Kepler field of view divides them in two groups \citep{nielsen2013}: O-A fast rotators (with periods of the axial rotation $P_{ax}\approx 2$ days), and slow rotators (G-M stars, $P_{ax}\approx 15$ days). A magnetic stellar wind is a possible breaker of the rotation of low mass main-sequence stars \citep{skumanich1972,iben1984}.

Observed rotational periods of WDs have a significant dispersion in the range $0.1-10$ days \citep{kawaler2015} with a usual value of the period about one day \citep{spruit1998}. It is close to the axial rotation period of massive main-sequence stars and about ten times shorter than the rotational period of low mass stars ($0.8-1.5M_{\odot}$). Observed rotational periods of the most young radio pulsars associated with supernova remnants are in the range $0.03-1$ seconds \citep{popov2012}. The rotation of stellar BHs in close binaries can approach to the critical value (e.g., \citealp{batta2017}), whereas rotational periods of single BHs remain unknown.

Rotational periods of main-sequence stars and their products in the forms of WDs and NSs can be compared. The aim of this comparison is to estimate the rotation rate and the acceleration efficiency of the core contraction in the course of stellar evolution with the braking efficiency of the stellar envelope expansion. We take here the core and the envelope as rigidly rotating bodies rotating independently. For the first approximation one can assume here (following the angular momentum conservation law) the rotational period

\begin{equation}
\label{p-rot:simple}
P_{rot}\sim R^2\sim \rho^{-\frac{2}{3}},
\end{equation}

\noindent where $R$ is the radius of the core that forms the remnant, $\rho$ is the average density of the core. Such simple method is now applied in quite unclear situation with the evolution of rotation of stellar nuclei. Thus, for the rotational period of WDs, if the stellar core conserves its angular momentum, we shall get $P_{min}\simeq 10^{-4}$ days if the main-sequence predecessor's mass was more than $\approx 1.5 M_{\odot}$ and $\sim 10^{-3}$ days if it is lower. Both values are low in comparison to observed values \citep{kawaler2015}, the expected rotation is too fast in such case.

WDs are forming by red supergiants with radii $\sim 10^3R_{\odot}$. If the white dwarf's core rotates as its extended envelope the rotational period of the dwarf should be $P_{max}=10^5$ days, it is evidently inappropriate value. So, the rotation of the contracting core of the star is decelerated by the expanding envelope to an intermediate rotational rate. It is interesting that observed values of periods of the axial rotation of degenerate dwarfs have practically the same order of magnitude as the ``logarithmic'' (``geometrical'') average

\begin{equation}
\label{log:average}
P_{obs}=\sqrt{P_{min}P_{max}}\approx 10\ \text{days}.
\end{equation}

The same estimations of rotational periods for NSs give $P_{min}=10^{-6}$ and $P_{max}=10^4$ days. The geometrical average for these limits is about 30 seconds, that is one order of magnitude longer than the observed values $0.03-0.5$ sec \citep{popov2012}. Similar estimations for stellar mass BHs produced by single stars and by components of wide binaries yield $P_{min}=10^{-6}$ sec. To estimate $P_{max}$ for BHs one should have in mind that stars with masses above $\sim 30 M_{\odot}$ can have intense stellar wind and their radii remain below several radii of main-sequence stars with the same masses. So, $P_{max}=10$ days, the geometrical average of the rotational period is $\simeq 1$ sec.

\begin{figure}
\begin{center}
\includegraphics[scale=0.55]{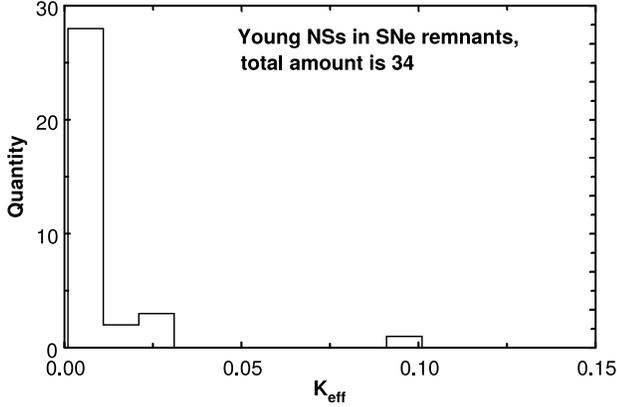}
\caption{A distribution of young NSs in supernovae remnants on the effective Kerr parameter according to Equation \ref{K_{eff}}. Initial spin periods of NSs were taken from Table 2 in \citet{popov2012}, if they were unknown we used current period values from the same table. Following assumptions were used to calculate $K_{eff}$: $k_I$=0.4 (a homogeneous ball), masses and radii of NSs were taken to be $1.4M_{\odot}$ and $10$ km respectively. }
\label{neutron}
\end{center}
\end{figure}

We can conclude that the acceleration of rotation of contracting nuclei is significantly broken by expanding envelopes, the observed rotational velocity of single WDs and NSs are far from limits $P_{min}$ and $P_{max}$ that we estimated above\footnote{For example, the helium core of the red-giant KIC 4448777 rotates almost rigidly and $\approx 6$ times faster than its convective envelope \citep{mauro2018}.}. Kerr parameters calculated for young NSs in supernova remnants (see Figure \ref{neutron}) are close to zero in most cases, it also tells us that the rotation of pre-supernova cores is far from the critical value. Close companions can significantly limit the evolutionary expansion of a star decreasing the braking efficiency of the expansion of envelopes of stars. This can be one of reasons of high rotational rates of close binary components. Red supergiants that precede the formation of degenerate dwarfs and NSs should have a strongly non-uniform rotation, their nuclei rotate many times faster than envelopes of such stars. Our treatment based on observational constraints is in some features similar to conclusions by \citet{aerts2018}.

\section{BH MASSES}
\label{sec:bh-masses}

The connection between the initial mass of a main-sequence star and the mass of its BH remnant is still uncertain, observational arguments are pure.
The model study of the evolution of massive stars depends on the input physics of uncertain mass loss, the final BH mass is also a function of the supernova explosion process, that is a function of several unknown parameters. As a first approximation we can admit here that the mass of a newborn BH is close to the mass of the carbon-oxygen core of a massive star at the moment of the collapse of its iron core. The collapse of the iron core of a massive ($25M_{\odot}-140M_{\odot}$) star is a result of the iron decomposition into alpha particles if the temperature of the stellar core exceeds $\sim 5\cdot 10^9$ K \citep{masevich1988,ohkubo2009}. The mass of the carbon-oxygen core according to numerical models \citep{masevich1988,heger2010} can be presented by the relation:

\begin{equation}
\label{equation:mass:carbon-oxygen}
\frac{M_{CO}}{M_{\odot}}=0.05\left(\frac{M_{init}}{M_{\odot}}\right)^{1.4},
\end{equation}

\noindent where $M_{init}$ is the initial mass of the main-sequence star. The mass of the helium remnant (a Wolf-Rayet star) of a massive component of a close binary is \citep{masevich1988}:

\begin{equation}
\frac{M_{He}}{M_{\odot}}=0.1\left(\frac{M_{init}}{M_{\odot}}\right)^{1.4},
\end{equation}

The minimal mass of a stellar mass BH according to Equation (\ref{equation:mass:carbon-oxygen}) is about $4.5 M_{\odot}$ if the initial main-sequence mass of the star was $25 M_{\odot}$. The maximum mass is about $55 M_{\odot}$ if the initial main-sequence mass of the star was $150 M_{\odot}$. This interval covers the observed interval of stellar mass BHs in X-ray binaries and in LIGO GW events. Combining Equation (\ref{equation:mass:carbon-oxygen}) and Salpeter mass function we can find that the mass function of stellar BHs is

\begin{equation}
\label{equation:bh-mass-distribution}
dN_{BH}=0.04\left(\frac{M_{\odot}}{M_{BH}}\right)^2\frac{dM_{BH}}{M_{\odot}}\ \textrm{per year}.
\end{equation}

The total frequency of BH formation in the Galaxy is about $0.007$ per year and the mass formation rate is about $0.015M_{\odot}$ per year. Therefore about 7\% of all matter that forms the stars finally should be transformed into stellar mass BHs. In general such distribution of BHs can be corrected if BHs may form from less massive stars than $25M_{\odot}$, and NSs are able to form from more massive stars \citep{sukhbold2016}, and in close binaries the accretion induced collapse can lead to the formation of a BH from a NS, see e. g. \citet{bogomazov2005a}. For these statements we do not take into account that there is a possibility for the existence of stars with masses higher than $150M_{\odot}$ \citep{popescu2014}. Also the mass distribution of stars can be different in comparison with the Salpeter mass function \citep{kroupa2001}.

\section{NECESSITY OF A COMMON ENVELOPE TO GET A COMPACT PRE-SUPERNOVA BINARY}
\label{sec:ce-nec}

Radius of massive main-sequence star can be presented by the relation \citep{masevich1988}:

\begin{equation}
\frac{R_{MS}}{R_{\odot}}\simeq \left(\frac{M_{He}}{M_{\odot}}\right)^{2/3},
\end{equation}

\noindent and the radius of a helium star by equation \citep{tutukov1973}:

\begin{equation}
\frac{R_{He}}{R_{\odot}}\simeq 0.2\left(\frac{M_{He}}{M_{\odot}}\right)^{0.6}.
\end{equation}

From Equation (\ref{equation:mass:carbon-oxygen}) follows that to explain the existence of a BH with mass $\approx 35 M_{\odot}$ in GW150914 \citep{abbott2016a} the initial mass of the star has to be $\sim 100 M_{\odot}$. But massive stars $M\geq 30 M_{\odot}$ potentially can quickly lose its matter by a wind and avoid expansion \citep{masevich1988,yusof2013,massey2017}. Thus, a common envelope stage that can move components close to each other was probably absent for progenitors of most massive BHs. Such binaries start from large separation and loose matter, they cannot become more close and would need too much time to merge. A possible way to explain the merger of a close pair of BHs is an assumption about a low metal abundance, which leads to decrease of their wind intensity, see e.g. \citet{belczynski2016}. Another possible way is a supposition about initially multiple system degrading during decay to a very close binary, e.g. \citet{antonini2016}. Nevertheless, \citet{bogomazov2018} showed that it is possible to produce a massive binary BH (that can merge within an appropriate interval of time) with or without the common envelope phase as the result of the isolated binary evolution.

\section{ROTATION OF COMPONENTS OF CLOSE BINARY BLACK HOLES}
\label{sec:bh-rotation}

We assume solid body rotation of pre-supernova synchronized with the orbital rotation at the moment of carbon-oxygen core formation following \citet{heuvel2007}, see also \citet{tutukov2003}. For collapsing cores of WR stars we calculate the value of ``effective Kerr parameter''

\begin{equation}
\label{K_{eff}}
K_{eff}=\frac{I\Omega}{GM^2_c/c};
\end{equation}

\noindent where $I=k_I M_c R_c^2$, $M_c$ is the mass of CO core, $R_c$ is its radius, $c$ is the speed of light, $k_I=0.4$ for a homogeneous ball, $k_I=0.1$ for a polytrope sphere with the polytropic index $n=2.5$.

The radius of the core after the end of helium burning can be found from the virial theorem:

\begin{equation}
\label{co:radius}
R_c=\frac{G\mu m_p M_c}{6kT};
\end{equation}

\noindent where $R_c$ and $M_c$ are the core radius and mass, $T$ is the temperature of carbon burning ($T\approx 6 \cdot 10^8$ K, \citealp{masevich1988}), $G$ is the gravitational constant, $\mu$ is the average number of nucleons for a particle (we take $\mu= 15$), $m_p$ is the proton mass, $k$ is the Boltzmann constant. 

After the formation the CO core evolves without significant angular momentum loss. This idea was used to explain long gamma ray bursts as supernova explosions in close binary systems with the orbital period less than about 1 day \citep{bogomazov2007,lipunov2007,bogomazov2009,lipunova2009}.

In this work we consider two cases of rotation of collapsing cores of progenitor of BH+BH mergers. The first one: all young CO cores are tidally locked independently on the orbital period. The second case: the core is tidally locked if the orbital period of the binary is $\leq 1$ day before the explosion of this core, otherwise we take a geometrical average axial period (described above in Section \ref{sec:remnant-rotation}) for the remnant of the exploding star. As an example of a geometrical average rotation let us consider a blue supergiant $\epsilon$ Ori (Alnilam). Its mass is $\approx 40M_{\odot}$, its radius is $\approx 32R_{\odot}$, rotational velocity $v\cdot \sin i=40\div 70$ km s$^{-1}$ \citep{searle2008,puebla2016}, from these quantities the derived period of axial rotation is about 23 days or higher. Equation \ref{equation:mass:carbon-oxygen} gives $M_{CO}=8.74M_{\odot}$, Equation \ref{co:radius} gives $R_c=3.87R_{\odot}$. For the axial period equal to 2 days for the main sequence progenitor of Alnilam (it is very common value for O-A stars, \citealp{nielsen2013}) we estimate $P_{min}=0.0004$ seconds (formally this rotation rate is faster than the rotation of a BH with Kerr parameter equal to 1), $P_{max}=23$ days, their geometrical average is $\approx 8.3$ seconds. For this period one can obtain $K_{eff}=0.01$ that we use in our calculations for stars in binaries with the orbital period longer than 1 day before the supernova explosion.

The described dependence of the angular momentum of the collapsing core is steep: core's rotation in closest pairs is very fast and these cores are able to leave a remnant with high spin, whereas in other binaries remnant's spin is low. In general, careful calculations of angular momentum evolution probably can be made \citep{postnov2017,qin2018}. But, if one take into account that the synchronization time depends on the star's radius $R$ and the major semi-axis $a$ as $(R/a)^6$ \citep{zahn1977,hut1981}, it can be concluded that for massive progenitors of BHs (their lifetime is very short) the synchronization of axial rotation with orbital rotation exists in closest binaries (and allows the formation of long gamma ray bursts during supernova explosions in them) and practically does not exist for wider pairs. All intermediate cases (partial synchronization) are able to exist, but the probability of their existence is very low due to the sixth power dependence of the timescale on the separation of the components.

We assume that the angular momentum vector of the axial rotation of collapsing cores and their remnants (BHs) aligns with the angular momentum vector of the orbital rotation. So, we compute the effective inspiral spin parameter (e.g. Equation 6 by \citealp{abbott2016a}) $\chi_{eff}$ using the expression

\begin{equation}
\label{khi-eff}
\chi_{eff}=\frac{M_{BH1}K_{eff1}+M_{BH2}K_{eff2}}{M_{BH1}+M_{BH2}},
\end{equation}

\noindent where $M_{BH1}$ and $M_{BH2}$ are the masses of two merging BHs taken from population synthesis modeling using the ``Scenario Machine'', $K_{eff1}$ and $K_{eff2}$ (see Equation \ref{K_{eff}}) are effective spins of collapsing cores that left merging BHs as remnants. To calculate $\chi_{eff}$ using Equation \ref{khi-eff} we assume $K_{eff}=1$ if $K_{eff}$ estimated using Equation \ref{K_{eff}} is $>1$, because the value of $\chi_{eff}$ belongs to merging BHs which spins cannot be higher than one.

\section{THE SCENARIO MACHINE}
\label{section-scm}

The ``Scenario Machine'' is a tool for population synthesis of the evolution of close binaries. It was introduced by \citet{kornilov1983}, developed by \citet{lipunov1996}, its latest detailed description can be found in a paper by \citet{lipunov2009}. In this paper we briefly describe only basic initial distributions and several parameters of binary evolution that are free in this investigation.

The following distribution of the initial semi-major axis $a$ is used:

\begin{equation}
\begin{cases}
\frac{dN}{d\left(\log a\right)} = 0.2,  \\
max\left(10 R_{\odot}, RL\left[M_1\right]\right)\le a \le 10^6 R_{\odot}.
\end{cases}
\label{dista}
\end{equation}

\noindent where $RL\left[M_1\right]$ is the size of the Roche lobe of the primary (initially more massive) star.

The initial mass $M_1$ of the primary is distributed using the Salpeter function:

\begin{equation}
\begin{cases}
f\left(M_1\right)=M_1^{-2.35},  \\
M_{min}\le M_1\le M_{max}.
\end{cases}
\label{distm}
\end{equation}

\noindent where $M_{min}$ and $M_{max}$ are the minimum and the maximum values of $M_1$ on the zero age main sequence.

The ``Scenario Machine'' contains three evolutionary scenarios: A, B and C. They differ in the rate of mass loss via the stellar wind of non-degenerate stars and in masses of cores of such stars.

In the present study we use four free parameters of stellar evolution: the rate of wind mass loss from stars, the fraction of the mass of the pre-supernova star that falls under the event horizon during the formation of a BH, the efficiency of the common envelope stage.

The stellar mass loss rate $\dot M$ is very important for two reasons: it significantly affects the semi-major axis of the binary and it directly affects the mass of the star itself. In the scenario A the mass loss by main-sequence stars is described here using the formula, see e. g. \citep{massevitch1979}:

\begin{equation}
\dot M=\frac{\alpha L}{cV_{\infty}}
\label{awind}
\end{equation}

\noindent where $L$ is the star's luminosity, $V_{\infty}$ is the stellar wind velocity at infinity, $c$ is the speed of light, and $\alpha$ is a free parameter. In scenario A, the decrease in the star's mass $\Delta M$ does not exceed 10\% of its hydrogen envelope during one evolutionary stage. We parametrized the mass loss by WR stars as

\begin{equation}
\Delta M_{WR}=\alpha_{WR}M_{WR},
\label{awindwr}
\end{equation}

\noindent here $M_{WR}$ is the initial stellar mass in the WR stage.

The mass of a BH $M_{BH}$ formed by an exploding presupernova of mass $M_{preSN}$, was calculated as

\begin{equation}
M_{BH}=k_{bh} M_{preSN},
\label{kbh}
\end{equation}

\noindent where the coefficient $k_{bh}$ is the fraction of the presupernova mass that forms the BH. 

During the common envelope stage binary stars very efficiently lose their angular momentum with the lost envelope matter, and the components approach one another along a spiral trajectory. The efficiency of mass loss in the common envelope stage is described by the parameter $\alpha_{CE}=\Delta E_b/\Delta E_{orb}$, where $\Delta E_b=E_{grav}-E_{thermal}$ is the binding energy of the ejected envelope and $\Delta E_{orb}$ is the decrease of the orbital energy during the approach:

\begin{equation}
\alpha_{CE}\left(\frac{GM_a M_c}{2a_f}-\frac{GM_a M_d}{2a_i}\right)=\frac{GM_d (M_d - M_c)}{R_d},
\label{ace}
\end{equation}

\noindent here $M_c$ is the core mass of the mass losing star with initial mass $M_d$ and radius $R_d$ (this is a function of the initial semi-major axis $a_i$ and initial component mass ratio $M_a/M_d$, where $M_a$ is the mass of the accretor), $a_f$ is the final semi-major axis in the end of the common envelope stage.

The common envelope forms if the Roche lobe overflow occurs\footnote{According to Kippenhahn and Weigert classification and Webbink diagram. See the description of the code by \citet{lipunov2009} for more details, subsection 4.4 ``RL Filling''.} in a type C system (where the star that fills its Roche lobe has a strongly evolved core), the whole mass ratio range, even for $q\sim 1$. For type B systems, we use the condition $q\le q_{cr} = 0.3$ for the formation of the common envelope, otherwise the systems evolves without the common envelope. \citet{heuvel2017}\footnote{See also \citep{pavlovskii2017}.} studied very similar conditions as for type B systems for the common envelope formation for the study of WR$+$O stars.

Evolutionary scenario B: several series of evolutionary tracks were calculated at the end of the 1980s and the beginning
of the 1990s, using new opacities \citep{rogers1991,kurucz1991}, nuclear reaction cross sections \citep{landre1990}, and parameters for convection in stars \citep{stothers1991}. For stars with masses
$M < 10M_{\odot}$, these tracks nearly coincide with those
calculated in scenario A. More massive stars should have
stronger stellar winds. A massive star loses up to 90\% of its mass due to stellar wind on the MS, supergiant and WR
stages. Therefore the mass of a pre-supernova could be $8-10M_{\odot}$ practically independently on the initial mass of the star.

Scenario C was based on the paper by \citet{vanbeveren1998}. The mass loss rate in the stellar wind was corrected to take into
account empirical data for OB and WR stars. The mass loss for stars with mass higher than $15M_{\odot}$ is parametrized by

\begin{equation}
\label{dm-c}
\Delta M = p_m(M - M_{core}),
\end{equation}

\noindent where $0<p_m\leq 1$ is a dimensionless parameter. Originally scenario C uses this equation without a parameter $p_m$ (i.e, $\Delta M = (M - M_{core})$), but for this work we introduced it to decrease the mass loss rate via the stellar wind, because without such correction it failed to explain the origin of most massive BH mergers \citep{bogomazov2018} and the existence of such X-ray binaries as IC 10 X-1 and NGC 300 X-1 \citep{abubekerov2009,bogomazov2014}.

Evolutionary scenarios B and C have several specific properties compared to the classical scenario. One
of them is the stronger stellar winds of massive stars lead to a rapid and substantial increase of the orbital separation, so these stars cannot fill their Roche lobes. There exist several observational facts that argue against scenarios with strong stellar winds (they can be found in \citealp{lipunov2009,bogomazov2018}), but we make calculations also for scenarios B and C to show that in a wide range of evolutionary parameters the isolated binary evolution can describe all detected gravitational wave bursts.

We assumed zero natal kick during the BH formation: observed BH+BH mergers favour low values of the BH natal kick \citep{wysocki2018}.

\section{CALCULATIONS AND RESULTS}

\begin{figure}[t!]
\begin{center}
\includegraphics[scale=1.87]{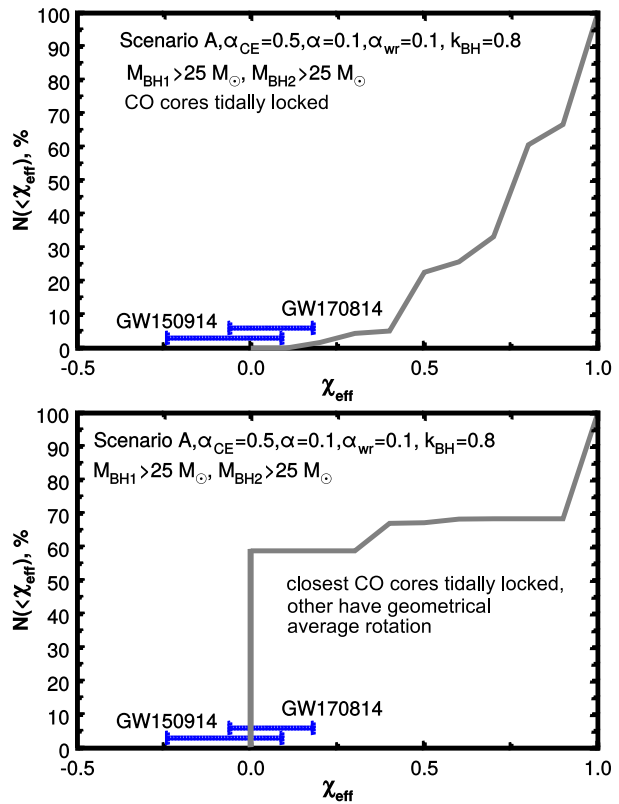}
\caption{ Cumulative distributions of of cores of future BH mergers on the $\chi_{eff}$ parameter as a part of all mergers with $\chi_{eff}$ equal or less than a definite value. The upper panel presents the assumption that the rotation of CO cores is tidally locked when such cores form (and cores conserve their angular momentum later) in all binaries that give BH+BH mergers in the end of their evolution. For the lower panel we assume that the axial rotation of the collapsing core is a geometrical average between the longest and shortest possible values if the orbital period $P_{orb}>1$ day before its explosions, if $P_{orb}\leq 1$ day before the core's explosion we assume that core is tidally locked as for the upper panel. For these curves we use following scenario parameters: evolutionary scenario A, $\alpha_{CE}=0.5$, $\alpha=\alpha_{WR}=0.1$, $k_{BH}=0.8$. Both masses of merging BH are greater than 25 solar masses. Observed GW events with such masses (see Table \ref{table1}) are shown with their error bars. }
\label{cumulative2a}
\end{center}
\end{figure}

\begin{figure}[t!]
\begin{center}
\includegraphics[scale=1.87]{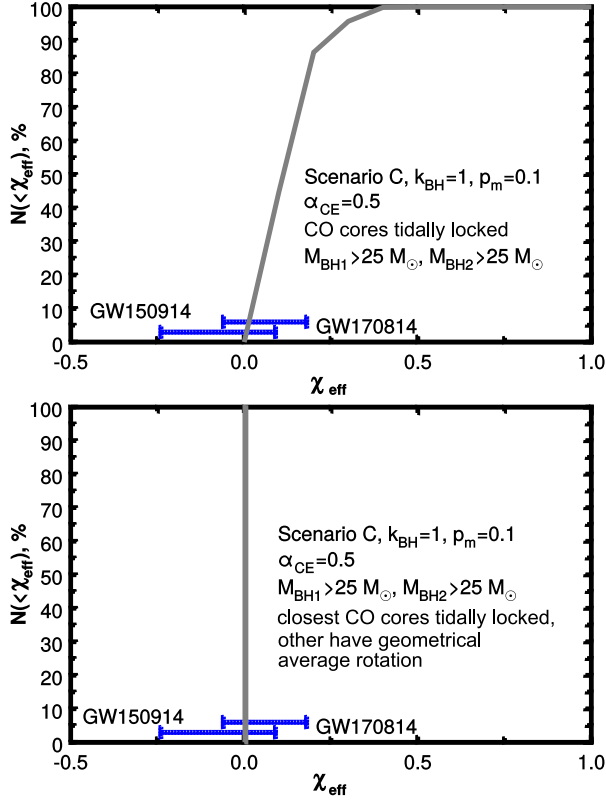}
\caption{ The same as Figure \ref{cumulative2a} for following scenario parameters: evolutionary scenario C, $k_{BH}=1$, $p_m=0.1$, $\alpha_{CE}=0.5$. }
\label{cumulative2c}
\end{center}
\end{figure}

\begin{figure}[t!]
\begin{center}
\includegraphics[scale=1.87]{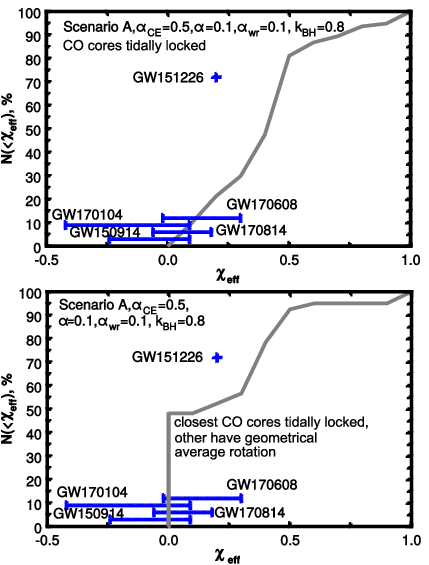}
\caption{ The same as Figure \ref{cumulative2a} for whole mass range of merging BHs. }
\label{cumulative-a}
\end{center}
\end{figure}

\begin{figure}[t!]
\begin{center}
\includegraphics[scale=1.87]{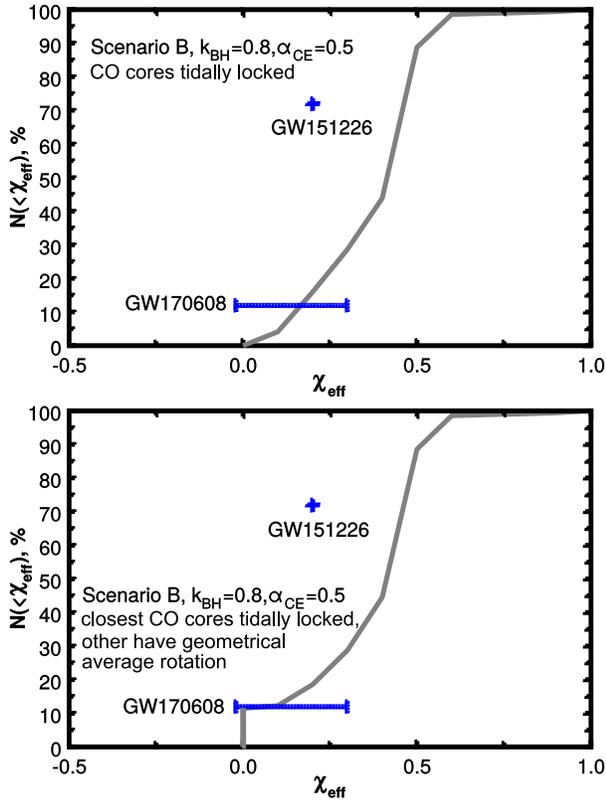}
\caption{ The same as Figure \ref{cumulative2a} for following scenario parameters: evolutionary scenario B, $k_{BH}=0.8$, $\alpha_{CE}=0.5$, whole mass range of merging BHs. In this Figure we show only two GW events with lowest masses, because in scenario B a BH with mass higher than about $10M_{\odot}$ cannot be produced. }
\label{cumulative-b}
\end{center}
\end{figure}

\begin{figure}[t!]
\begin{center}
\includegraphics[scale=1.87]{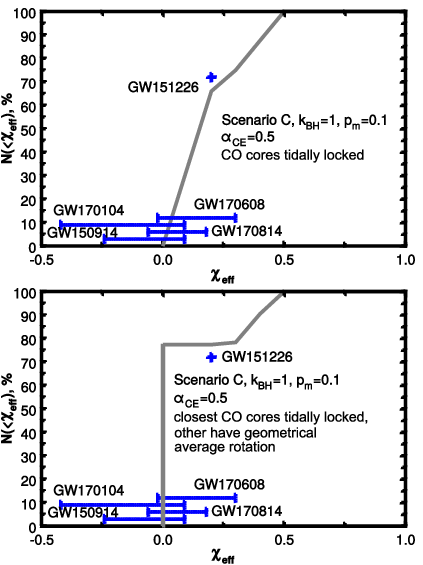}
\caption{ The same as Figure \ref{cumulative2c} for whole mass range of merging BHs. }
\label{cumulative-c}
\end{center}
\end{figure}

\begin{figure}[t!]
\begin{center}
\includegraphics[scale=1.87]{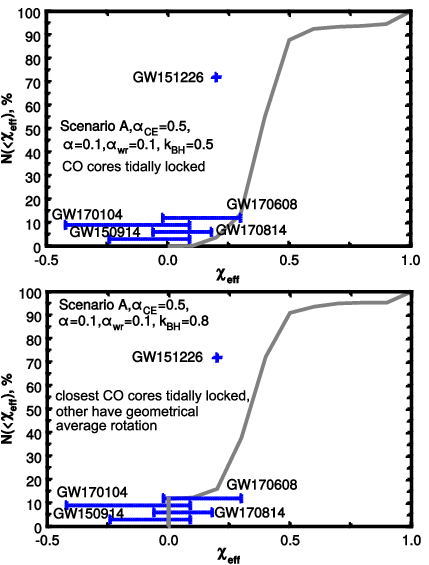}
\caption{ The same as Figure \ref{cumulative2a} for following scenario parameters: evolutionary scenario A, $\alpha_{CE}=0.5$, $\alpha=\alpha_{WR}=0.1$, $k_{BH}=0.5$, whole mass range of merging BHs. }
\label{cumulative-f-05-01}
\end{center}
\end{figure}

\begin{figure}[t!]
\begin{center}
\includegraphics[scale=1.87]{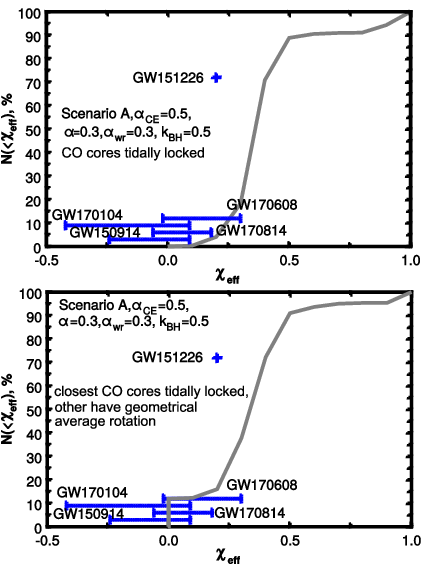}
\caption{ The same as Figure \ref{cumulative2a} for following scenario parameters: evolutionary scenario A, $\alpha_{CE}=0.5$, $\alpha=\alpha_{WR}=0.3$, $k_{BH}=0.5$, whole mass range of merging BHs. }
\label{cumulative-f-05-03}
\end{center}
\end{figure}

\begin{figure}[t!]
\begin{center}
\includegraphics[scale=1.87]{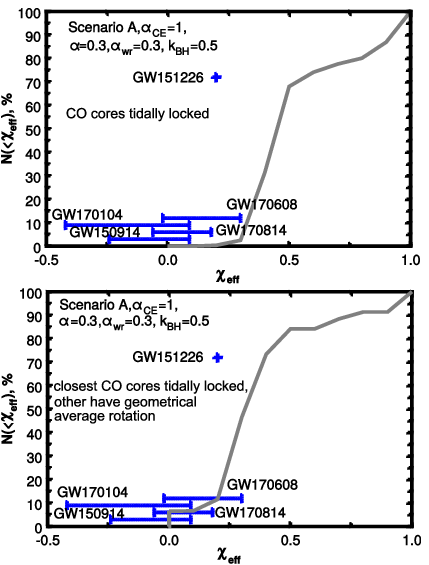}
\caption{ The same as Figure \ref{cumulative2a} for following scenario parameters: evolutionary scenario A, $\alpha_{CE}=1$, $\alpha=\alpha_{WR}=0.3$, $k_{BH}=0.5$, whole mass range of merging BHs. }
\label{cumulative-f-05-03-10}
\end{center}
\end{figure}

The ranges of possible values of scenario parameters are very wide, so in general all quantities used in population synthesis can be validated using the observational quantities of specific binaries, see e.g. such investigation for the ``Scenario Machine'' by \citet{lipunov1996}. Since the primary aim of this work is to show that binary evolution models are plausible to explain detected gravitational wave events we show here results of calculations for several models that were computed using following very different sets of free scenario parameters:

\begin{enumerate}

\item Models 1 and 1a. Scenario A, $\alpha_{CE}=0.5$, $\alpha=\alpha_{WR}=0.1$, $k_{BH}=0.8$, $M_{BH1,2}>25M_{\odot}$ (Figure \ref{cumulative2a}).

\item Models 2 and 2a. Scenario C, $k_{BH}=1$, $p_m=0.1$, $\alpha_{CE}=0.5$, $M_{BH1,2}>25M_{\odot}$ (Figure \ref{cumulative2c}).

\item Models 3 and 3a. Scenario A, $\alpha_{CE}=0.5$, $\alpha=\alpha_{WR}=0.1$, $k_{BH}=0.8$, whole mass range of merging BHs (Figure \ref{cumulative-a}).

\item Models 4 and 4a. Scenario B, $k_{BH}=0.8$, $\alpha_{CE}=0.5$, whole mass range of merging BHs (Figure \ref{cumulative-b}).

\item Models 5 and 5a. Scenario C, $k_{BH}=1$, $p_m=0.1$, $\alpha_{CE}=0.5$, whole mass range of merging BHs (Figure \ref{cumulative-c}).

\item Models 6 and 6a. Scenario A, $\alpha_{CE}=0.5$, $\alpha=\alpha_{WR}=0.1$, $k_{BH}=0.5$, whole mass range of merging BHs (Figure \ref{cumulative-f-05-01}).

\item Models 7 and 7a. Scenario A, $\alpha_{CE}=0.5$, $\alpha=\alpha_{WR}=0.3$, $k_{BH}=0.5$, whole mass range of merging BHs (Figure \ref{cumulative-f-05-03}).

\item Models 8 and 8a. Scenario A, $\alpha_{CE}=1$, $\alpha=\alpha_{WR}=0.3$, $k_{BH}=0.5$, whole mass range of merging BHs (Figure \ref{cumulative-f-05-03-10}).

\end{enumerate}

\noindent In this list in all models with mark ``a'' we assume that $K_{eff1,2}=0.01$ if the orbital period of the binary $P_{orb}> 1$ day before the explosion of the appropriate star, and if $P_{orb}\leq 1$ we accept that CO core in the beginning of its evolution is tidally locked and then it evolves without angular momentum loss (we calculate $K_{eff}$ using Equation \ref{K_{eff}}), in all model without mark ``a'' we assume only the last option (young CO cores are tidally locked with subsequent evolution without angular momentum loss) as an upper limit of core's rotation rate. For each model we calculated $10^6$ evolutionary tracks of individual binaries starting from the mass $25M_{\odot}$ in Equation \ref{distm} to avoid calculations of systems that do not produce BHs, then calculated values were renormalized for a galaxy with star formation rate as in our galaxy.

The results of our calculations are shown in Figures \ref{cumulative2a}--\ref{cumulative-f-05-03-10} and in Table \ref{table2}. In each figure the continuous line depicts a calculated cumulative distribution of BH mergers on $\chi_{eff}$ (see Equation \ref{khi-eff}) for a certain set of parameters. Detected mergers with appropriate mass are shown with their uncertainties in $\chi_{eff}$ (their vertical positions are chosen for graphical purposes only). As we can see from the figures all models with ``a'' mark describe observational points within their error bars, and models without ``a'' mark also seem to make it satisfactory.

In Table \ref{table2} we collected BH merge frequencies in all mentioned models, also we divided mergers into three groups in each model: ``high spin+high spin'' (effective Kerr parameter of the core $K_{eff1,2}\geq 1$), ``high spin+low spin'' ($K_{eff}$ equals or higher that 1 for one core and is less than 1 for another core), ``low spin+low spin'' ($K_{eff1,2}<1$). From this table we can see that all models produce ``low spin+low spin'' BHs that can be within the range of detected parameters of BH mergers. Examples of evolutionary tracks that lead to the formation of BHs with different spins are shown in Section \ref{appendix}.

\begin{table}[t!]
\caption{Individual BH spin distribution in BH mergers. Here ``Model'' is a set of evolutionary parameters (see text for details), ``L+L'' is the part of merging BHs with low spins of both merging BHs ($K_{eff}$ of both stars is $<1$), ``L+H'' is the part of low spin with high spin BHs ($K_{eff}$ is $<1$ for one star and is $\geq 1$ for another), and ``H+H'' is the part of two high spin BHs among all merging BHs ($K_{eff}$ of both stars is $\geq 1$), ``M. Fr.'' is the frequency of mergers in a galaxy like Milky Way. }
\begin{tabular*}{\columnwidth}{@{}l\x l\x l\x l\x l@{}}
\hline\hline
Model & H+H, \% & L+H, \% & L+L, \% & M. Fr., yr$^{-1}$ \\
\hline
1 & 33 & 54 & 13 & $9.6\cdot 10^{-6}$ \\
1a & 31 & 10 & 59 & $9.6\cdot 10^{-6}$ \\
2 & 0 & 0 & 100 & $8.4\cdot 10^{-5}$ \\
2a & 0 & 0 & 100 & $8.4\cdot 10^{-5}$ \\
3 & 5 & 50 & 45 & $1\cdot 10^{-4}$ \\
3a & 5 & 37 & 58 & $1\cdot 10^{-4}$ \\
4 & 1 & 52 & 47 & $1.4\cdot 10^{-4}$ \\
4a & 1 & 52 & 47 & $1.4\cdot 10^{-4}$  \\
5 & 0 & 20 & 80 & $1.2\cdot 10^{-4}$ \\
5a & 0 & 20 & 80 & $1.2\cdot 10^{-4}$ \\
6 & 5 & 76 & 19 & $1.1\cdot 10^{-4}$ \\
6a & 5 & 69 & 26 & $1.1\cdot 10^{-4}$ \\
7 & 5 & 77 & 18 & $9\cdot 10^{-5}$ \\
7a & 4 & 70 & 26 & $9\cdot 10^{-5}$ \\
8 & 13 & 82 & 5 & $8.5\cdot 10^{-5}$ \\
8a & 8 & 82 & 10 & $8.5\cdot 10^{-5}$ \\
\hline\hline
\end{tabular*}
\label{table2}
\end{table} 

\section{DISCUSSION}
\label{discussion}

The stellar wind is a substantial factor of the evolution of massive stars. Observations revealed that its intensity for OB stars can be described by the relation \citep{vink2018}:

\begin{equation}
\label{wind-z}
\dot{M}\simeq 10^{-9} (M_{OB}/M_{\odot})^2(Z/Z_{\odot})^{3/4}\ M_{\odot}\textrm{yr}^{-1};
\end{equation}

It indicates that stars with masses higher than $\approx 80(Z_{\odot}/Z)^{0.6}$ can lose their hydrogen envelopes during their main-sequence lifetime and probably can become Wolf-Rayet stars avoiding supergiant stage \citep{tutukov1973-2}. This factor can also be important for massive binary systems, because the abundance of metals should be adequately low to form a common envelope that can shrink the separation of the stars in the binary. Assuming $M_{BH}\simeq 0.05(M_{OB}/M_{\odot})^{1.4}$ \citep{masevich1988,nakauchi2018} one can find that the abundance of metals should meet following criterion to allow expanding of components of the binary and to allow the common envelope formation:

\begin{equation}
\label{inmet}
Z\lesssim (25M_{\odot}/M_{BH})^{1.2}Z_{\odot};
\end{equation}

If this condition is met very close massive binary BHs can form and they can merge under the influence of gravitational wave radiation during Hubble time. Therefore predecessor of the most massive merging BHs should have low metal abundance \citep{tutukov2017}.

Low metal abundance (that is common for first stars in the Universe) leads to an important consequence for estimations of merge frequency of BHs. The decrease of the metal abundance to [Fe/H]$\lesssim -0.1$ leads to the growth of power of the initial mass function from $\approx -2.5$ for star with solar metallicity to $\approx -2$ \citep{sarzi2018}. Evidently the quantity of stars that form high mass BHs $\gtrsim 30 M_{\odot}$ grows by 5-10 times, and the part of the most massive BHs among all merging BHs also grows.

So, the decrease of metal abundance in early Universe is able to make the mass specturm of merging BHs wider due to saving the possibility to form a common envelope, and also it increases the quantity of massive stars due to the change of the initial mass function. This probably is an explanation of high masses of $>80$\% of detected merging BHs in comparison with their part in X-ray binaries ($\approx 13$\%,  \citealp{mapelli2018}).

It is essential to note that observational data about the distribution of eccentricities of orbits of WR+OB binaries with solar metallicity in our galaxy \citet{cherepashchuk2018}, and axial rotation velocities of O stars in such systems \citet{shara2017, vanbeveren2018} indicate that the mass exchange in most of known WR+OB binary stars during WR stars formation is the primary cause of the envelope loss, whereas radial stellar wind is only the secondary cause. This fact can be related to wind clamping and observational overestimates of mass loss rate via the stellar wind \citep{cherepashchuk1990}. Wind clumps were observed by e.g. \citet{markova2004, markova2005, puls2006, markova2008, furst2010, ducci2009, oskinova2012}.

So, even for massive stars with near solar metallicity the classical scenario of isolated binary evolution with mass exchange can be considered as very probable mechanism of the formation of close pairs BH+BH, BH+NS, NS+NS that subsequently merge and form gravitational wave events.  That is why we can treat evolutionary scenario A as the primary evolutionary scenario that describes the evolution of close binary stars adequately. The main cause of the absence of very massive BH candidates in X-ray binaries in observations and in theoretical predictions (e.g. made by \citealp{fryer2001,bogomazov2005b}) is the very short evolutionary time scale of the most massive stars. The common envelope formation condition described in Section \ref{section-scm} plays very important role. It allows to get systems that goes through the common envelope before the first supernova explosion in the system, or before the second explosion, or even to get a system that does not go through the common envelope at all, so the axial rotation rate of the pre-supenova star can vary from low to high rate depending on the evolutionary way of the system \citep{bogomazov2018}.

Our calculations show (Figures \ref{cumulative2a}-\ref{cumulative-f-05-03-10} and Table \ref{table2}) that a part of collapsing core in massive close binaries possess supercritical angular momentum. Such collapsing cores are expected to be sources of long gamma ray burst, but at the same time the collapse itself can consist of two stages. During the first stage a binary BH forms, the separation of its components can be several radii of BHs. Due to the emission of gravitational waves by such massive and tight BHs the binary BH should merge in time less than 1 second. So, two tasks arise. The first is to clarify the condition of the origin of supercritical rotation of collapsing cores that form BHs. The second is to calculate sample signals of gravitational waves from the two stage collapse in order to unambiguously identify them experimentally.

\section{CONCLUSIONS}

A population synthesis using wide range of evolutionary parameters was conducted. Our calculations allow to make following conclusions:

\begin{itemize}

\item Isolated evolution of binary stars adequately describes detected gravitational wave events.

\item Different values of spins of BH components in BH mergers (low spin + low spin, low spin + high spin, high spin + high spin) are able to exist. Individual BHs in future BH mergers should rather have low value of spin (close to zero) or high value of spin (close to one) because of sixth power dependence of the syncronization time on the separation of their non-degenerate predecessors and a short evolution time scale for massive stars. Intermediate values are able to exist, but their birth probability is small.

\item $\geq 50$\% BH mergers originated through isolated binary evolution should possess spins of both BHs close to zero (low spin + low spin).

\item Up to  $\approx$10\% of BH mergers are able consist of two black holes with spins close to one (high spin + high spin), predecessors of such binaries could be sources of two gamma ray bursts.

\item The BH with the spin close to one could be formed as the result of the merger of two black holes formed via the collapse of a fast rotating star accompanied with a long gamma ray burst and/or a gravitational wave burst with smaller power in comparison with the merger of two BH remnants of binary components. Probably this idea can be tested in the near future \citep{nathanail2018, atteia2018}.

\end{itemize}

Our conclusions are very similar to conclusions by \citet{qin2018}, but we have different arguments in favour of the idea that at least 50\% of merging black holes should possess low spin values of both BHs. Also we predict that significant fraction of merging black holes should have at least one BH or even two BHs with spin close to 1 since they can be remnants of gamma ray bursts. This claim can be very important if individual spins of BH mergers can be found more or less reliably, because $chi_{eff}$ is a  smoothed function of $K_{eff1,2}$. The physic of collapse is not in the scope of this paper, nevertheless, it is very important for understanding described possibilities for BH spin values.

\begin{acknowledgements}

The work was supported by a grants of Russian Science Foundation: RSF 16-12-00085 (V. M. Lipunov; an analysis of spin values of neutron stars in supernova remnants) and RSF 17-12-01241 (A. M. Cherepashchuk; an analysis of the role of mass exchange in the evolution of WR+O binary stars).

\end{acknowledgements}

\section{APPENDIX. EXAMPLES OF EVOLUTIONARY TRACKS}
\label{appendix}

In Tables \ref{track-a-hh}-\ref{track-c-ll} we present examples of evolutionary tracks that lead to a formation of black holes with different spin values  (``high spin+high spin'', ``high spin+low spin'', ``low spin+low spin'') calculated using different sets of scenario parameters.

Notations in tables are following: ``I'' is the main-sequence star, ``II'' is the supergiant star, ``3'' is the star filling its Roche lobe, ``3E'' is the star filling its Roche lobe in evolutionary time scale, ``3S'' is the star filling its Roche lobe with supercritical accretion onto the compact companion, ``WR'' is the Wolf-Rayet star, ``CE'' is the common envelope, ``BH'' is the black hole, ``SBH'' is the black hole with supercritical accretion rate. Parameters of stars and parameters of binaries listed in tables are following: ``Stage'' is the evolutionary status of the system, ``$M_1$'' is the mass of the primary (initially more massive) star in $M_{\odot}$, ``$M_2$'' is the mass of the secondary star in $M_{\odot}$, ``$a$'' is the semi-major axis of the binary's orbit in $R_{\odot}$, ``$e$'' is its eccentricity, $T$ is the time elapsed after the beginning of the evolution in $10^6$ years.

\begin{table}[h!]
\caption{Evolutionary track that produces ``high spin+high spin'' BHs before merger. Following set of scenario parameters was used: scenario A, $\alpha_{CE}=0.5$, $\alpha=\alpha_{WR}=0.3$, $k_{BH}=0.5$. Notations are described in the text.}
\begin{tabular*}{\columnwidth}{@{}l\x l\x l\x l\x l\x l@{}}
\hline\hline
Stage & $M_1$ & $M_2$ & $a$ & $e$ & $T$ \\
\hline
I+I & 95.44 & 20.90 & 190 & 0  &  0 \\
3+I, CE & 90.18 & 20.43 & 200 & 0 & 2.5 \\
WR+3E & 59.10 & 31.09 & 10 & 0 & 2.5 \\
SN Ib \\
BH+3E & 31.00 & 27.49 & 25 & 0.53 & 2.7 \\
BH+WR & 31.00 & 27.49 & 12 & 0 & 3.5 \\
SN Ib \\
BH+BH & 31.00 & 9.62 & 18 & 0.24 & 3.9 \\
 Coalescence        
BH & 40.62 & & & & 880 \\
\hline\hline
\end{tabular*}
\label{track-a-hh}
\end{table}

\begin{table}[h!]
\caption{Evolutionary track that produces ``high spin+low spin'' BHs before merger. Following set of scenario parameters was used: scenario A, $\alpha_{CE}=0.5$, $\alpha=\alpha_{WR}=0.3$, $k_{BH}=0.5$. Notations are described in the text.}
\begin{tabular*}{\columnwidth}{@{}l\x l\x l\x l\x l\x l@{}}
\hline\hline
Stage & $M_1$ & $M_2$ & $a$ & $e$ & $T$ \\
\hline  
I+I & 75.07 & 50.02 & 140 & 0 & 0 \\
3+I & 72.08 & 48.95 & 140 & 0 & 2.6 \\
3E+I & 52.66 & 52.66 & 140 & 0 & 2.6 \\
WR+I & 42.23 & 55.64 & 160 & 0 & 3.2 \\
SN Ib \\
BH+I & 14.78 & 55.52 & 240 & 0.21 & 3.5 \\
BH+II & 14.78 & 55.41 & 240 & 0.21 & 3.7 \\
SH+3S & 14.78 & 50.76 & 240 & 0.1 & 4 \\
SH+3, CE & 14.78 & 45.58 & 160 & 0.09 & 4 \\
BH+WR & 14.78 & 27.77 & 8.2 & 0 & 4 \\
SN Ib \\
BH+BH & 14.78 & 9.72 & 17 & 0.4 & 4.3 \\
Coalescence \\
BH & 24.50 & & & & $1.7\cdot 10^{3}$ \\   
\hline\hline
\end{tabular*}
\label{track-a-hl}
\end{table}

\begin{table}[h!]
\caption{Evolutionary track that produces ``low spin+low spin'' BHs before merger. Following set of scenario parameters was used: scenario A, $\alpha_{CE}=0.5$, $\alpha=\alpha_{WR}=0.3$, $k_{BH}=0.5$. Notations are described in the text.}
\begin{tabular*}{\columnwidth}{@{}l\x l\x l\x l\x l\x l@{}}
\hline\hline
Stage & $M_1$ & $M_2$ & $a$ & $e$ & $T$ \\
\hline  
I+I & 114.05 & 32.90 & 250 & 0 & 0 \\
3+I, CE & 105.97 & 31.03 & 270 & 0 & 2.4 \\
WR+3E & 75.85 & 61.15 & 21 & 0 & 2.4 \\
 SN Ib                \\
SH+3 & 41.44 & 54.11 & 40 & 0.44 & 2.6 \\
SH+3E & 41.45 & 41.45 & 28 & 0.19 & 2.6 \\
BH+WR & 41.70 & 31.69 & 27 & 0 & 3.6 \\
 SN Ib                \\
BH+BH & 41.70 & 11.09 & 40 & 0.21 & 3.8 \\
 Coalescence          \\
BH & 52.79 & & & & $1.2\cdot 10^4$ \\  
\hline\hline
\end{tabular*}
\label{track-a-ll}
\end{table} 

\begin{table}[h!]
\caption{Evolutionary track that produces ``high spin+high spin'' BHs before merger. Following set of scenario parameters was used: scenario B, $\alpha_{CE}=0.5$, $k_{BH}=0.8$. Notations are described in the text.}
\begin{tabular*}{\columnwidth}{@{}l\x l\x l\x l\x l\x l@{}}
\hline\hline
Stage & $M_1$ & $M_2$ & $a$ & $e$ & $T$ \\
\hline  
I+I & 113.70 & 107.82 & 70 & 0 & 0 \\
I+3E &  37.80 & 35.96 & 210 & 0 & 2.6 \\
II+3E &  37.85 & 35.67 & 210 & 0 & 2.6 \\
3+3e, CE &  49.29 & 21.66 & 310 & 0 & 2.9 \\
WR+WR &   8.01 & 8.02 & 2.8 & 0 & 2.9 \\
 SN Ib                 \\
BH+WR &   4.49 & 5.71 & 4.5 & 0.11 & 3.1 \\
 SN Ib                 \\
BH+BH &   4.49 & 4.51 & 5.4 & 0.25 & 3.1 \\
 Coalescence           \\
BH &   & 9 & & & 490 \\
\hline\hline
\end{tabular*}
\label{track-b-hh}
\end{table}      

\begin{table}[h!]
\caption{Evolutionary track that produces ``high spin+low spin'' BHs before merger. Following set of scenario parameters was used: scenario B, $k_{BH}=0.8$. Notations are described in the text.}
\begin{tabular*}{\columnwidth}{@{}l\x l\x l\x l\x l\x l@{}}
\hline\hline
Stage & $M_1$ & $M_2$ & $a$ & $e$ & $T$ \\
\hline 
3E+I & 17.24 & 18.77 & 44 & 0 & 3.7 \\
WR+I &  8.67 & 23.54 & 86 & 0 & 4.7 \\
WR+II &  6.54 & 22.97 & 94 & 0 & 5.0 \\
 SN Ib               \\
BH+II &  4.95 & 22.77 & 100 & 0.05 & 5.0 \\
SH+3S &  4.95 & 21.23 & 110 & 0 & 5.6 \\
SH+3 &  4.95 & 19.26 & 63 & 0 & 5.6 \\
BH+WR &  4.95 & 10.73 & 2.4 & 0 & 5.6 \\
 SN Ib               \\
BH+BH &  4.95  & 6.01 & 3.5 & 0.14 & 6.1 \\
 Coalescence         \\
BH &   & 10.96 &  &  & 61 \\
\hline\hline
\end{tabular*}
\label{track-b-hl}
\end{table}   

\begin{table}[h!]
\caption{Evolutionary track that produces ``low spin+low spin'' BHs before merger. Following set of scenario parameters was used: scenario B, $k_{BH}=0.8$. Notations are described in the text.}
\begin{tabular*}{\columnwidth}{@{}l\x l\x l\x l\x l\x l@{}}
\hline\hline
Stage & $M_1$ & $M_2$ & $a$ & $e$ & $T$ \\
\hline 
I+I & 49.75 & 37.11 & 61 & 0 & 0 \\
3E+I & 19.78 & 21.27 & 130 & 0 & 3.2 \\
3E+II & 9.63 & 23.30 & 270 & 0 & 3.9 \\
WR+II & 8.12 & 24.81 & 340 & 0 & 4.0 \\
 SN Ib               \\
BH+II & 4.55 & 22.91 & 410 & 0.04 & 4.3 \\
SH+3S & 4.55 & 22.49 & 410 & 0.04 & 4.4 \\
SH+3, CE & 4.55 & 20.59 & 230 & 0.04 & 4.4 \\
BH+WR & 4.55 & 10.32 & 5.9 & 0 & 4.4 \\
 SN Ib               \\
BH+BH & 4.55 & 5.78 & 8.7 & 0.14 & 4.8 \\
 Coalescence         \\
BH & & 10.33 & & & $2.7\cdot 10^3$ \\
\hline\hline
\end{tabular*}
\label{track-b-ll}
\end{table}

\begin{table}[h!]
\caption{Evolutionary track that produces ``high spin+low spin'' BHs before merger. Following set of scenario parameters was used: scenario C, $p_m=0.1$, $k_{BH}=1$. Notations are described in the text.}
\begin{tabular*}{\columnwidth}{@{}l\x l\x l\x l\x l\x l@{}}
\hline\hline
Stage & $M_1$ & $M_2$ & $a$ & $e$ & $T$ \\
\hline
I+I & 31.15 & 25.56 & 5.3E+01 & 0 & 0 \\
3+I & 29.26 & 24.47 & 5.6E+01 & 0 & 5.6 \\
3E+I & 25.72 & 25.72 & 5.7E+01 & 0 & 5.6 \\
WR+I & 12.32 & 31.65 & 1.1E+02 & 0 & 7.0 \\
 Silent Collapse                        \\
BH+I & 12.07 & 31.02 & 1.2E+02 & 0 & 7.5  \\
BH+II & 12.07 & 29.18 & 1.2E+02 & 0 & 14  \\
SH+3S & 12.07 & 27.52 & 1.3E+02 & 0 & 14   \\
SH+3, CE & 12.07 & 25.71 & 1.1E+02 & 0 & 14 \\
BH+WR & 12.07 & 12.60 & 5.1E+00 & 0 & 14 \\
 Silent Collapse                        \\
BH+BH & 12.07 & 12.33 & 5.2E+00 & 0 & 15 \\
 Coalescence         \\
BH & & 24.41 & & & 40 \\
\hline\hline
\end{tabular*}
\label{track-c-hl}
\end{table}

\begin{table}[h!]
\caption{Evolutionary track that produces ``low spin+low spin'' BHs before merger. Following set of scenario parameters was used: scenario C, $p_m=01$, $k_{BH}=1$. Notations are described in the text.}
\begin{tabular*}{\columnwidth}{@{}l\x l\x l\x l\x l\x l@{}}
\hline\hline
Stage & $M_1$ & $M_2$ & $a$ & $e$ & $T$ \\
\hline
I+I & 47.61 & 35.78 & 37 & 0 & 0 \\
II+I & 46.85 & 35.45 & 37 & 0 & 4.8 \\
WR+I & 44.65 & 35.37 & 38 & 0 & 5.1 \\
 Silent Collapse  \\
BH+I & 41.32 & 35.26 & 40 & 0 & 5.4 \\
BH+II & 41.32 & 34.89 & 40 & 0 & 6.2 \\
 Silent Collapse  \\
BH+BH & 41.32 & 33.16 & 41 & 0 & 6.6 \\
 Coalescence \\
BH & 74.48 &  &  & & $3.5\cdot 10^3$ \\
\hline\hline
\end{tabular*}
\label{track-c-ll}
\end{table}

\bibliographystyle{pasa-mnras}
\bibliography{bogomazov}

\end{document}